%% file: paper.tex
\documentclass[usenatbib,letterpaper]{mn2e}
\usepackage[totalwidth=480pt, totalheight=680pt]{geometry}
\usepackage{times}
\usepackage{epsfig}
\usepackage{amssymb}
\usepackage{amsmath}
\usepackage{aas_macros}
\title[Satellite quenching]{Satellite quenching timescales in clusters from projected phase space measurements matched to simulated orbits}
\author[K. A. Oman \& M. J. Hudson]{\newauthor Kyle A. Oman$^{1,2}$\thanks{koman@uvic.ca}, Michael J. Hudson$^{2,3}$
\\
$^{1}$ Department of Physics and Astronomy, University of Victoria, Victoria, British Columbia, V8P 5C2, Canada\\
$^{2}$ Department of Physics and Astronomy, University of Waterloo, Waterloo, Ontario, N2L 3G1, Canada\\
$^{3}$ Perimeter Institute for Theoretical Physics, Waterloo, Ontario, N2L 2Y5, Canada\\
}
\date{\today}

\def\Msun{\hbox{$\rm\thinspace M_{\odot}$}}

\begin{document}
\label{firstpage}
\maketitle

\begin{abstract}
We measure the star formation quenching efficiency and timescale in cluster environments. Our method uses N-body simulations to estimate the probability distribution of possible orbits for a sample of observed SDSS galaxies in and around clusters based on their position and velocity offsets from their host cluster. We study the relationship between their star formation rates and their likely orbital histories via a simple model in which star formation is quenched once a delay time after infall has elapsed. Our orbit library method is designed to isolate the environmental effect on the star formation rate due to a galaxy's present-day host cluster from `pre-processing' in previous group hosts. We find that quenching of satellite galaxies of all stellar masses in our sample ($10^{9}-10^{11.5} \Msun$) by massive ($> 10^{13} \Msun$) clusters is essentially $100$~per~cent efficient. Our fits show that all galaxies quench on their first infall, approximately at or within a Gyr of their first pericentric passage. There is little variation in the onset of quenching from galaxy-to-galaxy: the spread in this time is at most $\sim 2$~Gyr at fixed $M_*$. Higher mass satellites quench earlier, with very little dependence on host cluster mass in the range probed by our sample.
\end{abstract}


\section{Introduction}\label{sec-introduction}

A detailed understanding of the mechanisms that quench star formation in galaxies remains elusive. It now seems clear that quenching is strongly correlated with an `internal' parameter that is closely related to galaxy mass: stellar mass \citep{2003MNRAS.341...33K, 2004ApJ...600..681B}, velocity dispersion \citep{SmiLucHud09c,GraFabSch09}, or structural properties, such as the central stellar surface mass density \citep{CheFabKoo12, FanFabKoo13}  or the bulge fraction \citep{OmaBalPog14, 2014MNRAS.441..599B}. The physical cause of this quenching is still not known, although AGN \citep{2004ApJ...600..580G, 2006MNRAS.370..645B, 2006MNRAS.365...11C} and/or mergers \citep{2006ApJS..163....1H} and/or disc instabilities \citep{2009ApJ...703..785D} are often cited. It has become clear that environment also plays a role: once a galaxy falls into a more massive halo (such as a group or cluster) and becomes a satellite, there is an additional probability of quenching over and above the stronger `mass-related' quenching \citep{BalBalNic04, 2008MNRAS.387...79V, 2010ApJ...721..193P}. It is this latter `satellite quenching' that is the subject of this paper.

An infalling, actively star-forming satellite galaxy might be affected by its host halo in several ways, as reviewed by \citet{BosGav06}. Ram pressure stripping may remove cold gas from the disc \citep{1972ApJ...176....1G}. It has long been known that cluster galaxies are HI-deficient \citep[e.g.][]{GioHay85}. Furthermore, ram pressure stripping has been observed in individual infalling galaxies in a number of nearby clusters. In particular, in the Coma cluster at least $40$~per~cent of blue galaxies within $500$~kpc of the centre have young stars formed from stripped material visible at ultraviolet wavelengths \citep{2010MNRAS.408.1417S}, which suggests that ram pressure stripping is ubiquitous. However, from these snapshots it is difficult to tell how rapid this process is, and how effective it is overall.  A closely-related physical process is `strangulation' \citep{1980ApJ...237..692L, 2000ApJ...540..113B} in which the hot gas halo is stripped by ram pressure, thus removing the source that would otherwise have replenished the cold gas in the disc. Because the cold gas is not immediately affected, the timescale for strangulation should be longer than ram pressure stripping of the cold gas disc.

One way to constrain the quenching mechanism(s) is by measuring how effective quenching is, where or when it first occurs in the satellite's orbit, and how long a satellite takes to quench. Environment has a role in regulating star formation on a range of host mass scales \citep[e.g.][]{2010ApJ...721..193P}. In this paper we focus on galaxy clusters. These lend themselves well to our methodology for two reasons. First because their centers and extents, in the sense of both position and velocity, are much better defined than in poorer galaxy groups (though centering becomes easier again for systems such as the Milky Way and its satellites). Second, our choice to study relatively large satellites around massive clusters means we are able to draw our sample of observed galaxies from a large volume, yielding a statistically powerful data set. While clusters host a relatively small fraction of the passive galaxy population -- only about 15~per~cent (2.5~per~cent) of red galaxies with $M_* \geq 10^{9}$\Msun are satellites in $\geq 10^{13}$\Msun ($10^{14}$\Msun) haloes (estimated using the galaxy stellar mass function of red galaxies from \citealp{2012MNRAS.421..621B} and the satellite fraction and host halo mass distribution of \citealp{2008MNRAS.387...79V}) -- they offer a useful proving ground for analysis techniques aimed at constraining the timescale(s) of the quenching process(es) before attempting to tackle the more difficult galaxy group scale. In addition, it may be that a single environmental quenching mechanism is dominant in host haloes of all masses (see for instance \citealp{2008MNRAS.387...79V}, though \citealp{2015arXiv150306803F} argue the opposite). In this case the study of quenching in clusters can directly inform more difficult studies of lower mass hosts.

Some early work involved comparing semi-analytic models to observations. In models where the quenching occurs quickly after crossing the host halo's virial radius, too many red satellites were produced. The disagreement suggests that quenching process had to be slow \citep{WeivanYan06a, 2008MNRAS.389.1619F, BalMcGWil09, WeiKauvon09, 2009MNRAS.394.1131K}. \citet{2010MNRAS.409..405H} showed that, while bulge colours do not depend on cluster-centric radius, the colours of discs are redder closer to the cluster centre. These results were modelled by \citet{TarHudBal14}, who found that star formation in discs declines with an exponential timescale of $\sim 3$~Gyr, starting at cluster infall.

\citet[hereafter W13]{2013MNRAS.432..336W} concluded that satellites are quenched on timescale of 2-6 Gyr after passing the virial radius of a larger host halo for the first time. W13 obtained this result by measuring quenched fractions of satellites and centrals in low-redshift SDSS groups and clusters, and comparing these data with a satellite quenching model based on a halo infall time distribution from N-body simulations combined with an empirically-calibrated model of the quenched fractions at higher redshifts. Similar results to those of W13 were obtained by \citet{2014MNRAS.444.2938H}. At higher redshifts ($z \sim 1$), \citet{2014MNRAS.438.3070M} found shorter timescales of order $1$~Gyr.

An alternative approach is to take advantage of galaxies' positions in the observational projected phase space (PPS) of separation in the plane of the sky and line of sight velocity. \citet{2005MNRAS.356.1327G} showed that, at the same projected radius, galaxies in different phases of their orbits have different kinematics. This was extended by \citet{2013MNRAS.431.2307O} who constructed a subhalo orbit library that allowed them to construct a detailed probabilistic mapping between position in PPS and subhalo infall time.

\citet{2011MNRAS.416.2882M} were the first to deproject the PPS to obtain constraints on star formation histories of galaxies falling into larger systems. They studied galaxies with recent (within $1-3$~Gyr) or ongoing star formation, and concluded that star formation is efficiently quenched in a single passage through the cluster. Other authors have used the PPS to understand quenching at high redshift \citep{2014ApJ...796...65M} or the effects of ram pressure stripping \citep{2014MNRAS.438.2186H,2015MNRAS.448.1715J}.

The aim of this paper is to study the star-formation rates of galaxies based on their location in PPS, and model these using the orbit libraries of \citet{2013MNRAS.431.2307O}. Whereas \citet{2011MNRAS.416.2882M} used a coarse binning of the populations (`virial', `backsplash', `infalling'), in this paper we use detailed orbit libraries drawn from N-body simulations. Our model consists of two components: (1) an infalling population of galaxies (which are observed predominantly outside the virial radius and are assumed to have some `pre-processed' quenched fraction) and (2) a simple model for quenching in which some fraction of the active infalling galaxies are quenched following a delay $\Delta t$ after passing $2.5\,r_{\rm vir}$. The model then predicts the quenched fraction at any point in PPS. The infalling quenched fraction is fit from the PPS data simultaneously with the free parameters of the model (the efficiency of quenching and the timescale of quenching). This allows us to account for `pre-processing' in a natural way, and hence our results isolate the physical effects of infall of active satellites into their \emph{current} cluster-mass ($\sim 10^{14.5} \Msun$) host haloes. This differs from the approach of W13, in which the quenching timescale refers to the time since infall into \emph{any} halo, and so includes processing in the current host halo plus `pre-processing' in host haloes of lower mass.

This paper is structured as follows: in \S\ref{sec-data} we describe our numerical and observed data samples. In \S\ref{sec-method} we describe our models and fitting method. In \S\ref{sec-results} we present the results of fitting our models to the observed data. We discuss our results and compare to other work in \S\ref{sec-conclusions} and summarize in \S\ref{sec-summary}.

We assume the same cosmology used in the Bolshoi and Multidark Run 1 simulations with $\Omega_m=0.27$, $\Omega_\Lambda=0.73$, $\Omega_b=0.0469$, $n_s=0.95$, $h_0=0.70$, $\sigma_8=0.82$ \citep{2012MNRAS.423.3018P}.

\section{Data}\label{sec-data}

\subsection{Numerical simulations}\label{subsec-simulations}

We use the output of the Multidark Run 1 (MDR1) dark matter-only cosmological simulation to obtain a large sample of satellite orbits. The simulation has a box side length of $1\,h^{-1}\,{\rm Gpc}$, $2048^3$ particles, $8.63\times10^{9}\,h^{-1} \Msun$ mass resolution, $7\,h^{-1}\,{\rm kpc}$ force resolution, and uses the \emph{WMAP} 7 cosmology. The simulation runs from redshift $z=65$ to $0$ and has outputs linearly spaced\footnote{The resolution in scale factor doubles after $a\sim0.7$. There are also a handful of irregularly spaced steps at small $a$.} in scale factor $a$. The time resolution at $z=0$ is of about $0.21$~Gyr. For further details regarding MDR1 we refer to \citet{2012MNRAS.423.3018P}. The simulation output was processed with the {\sc rockstar} halo finder \citep{2013ApJ...762..109B} and the merger tree code of \citet{2013ApJ...763...18B}. In order to use host-satellite linking in the merger tree as a proxy for cluster membership out to the largest apocentric radii of about $2.5\,r_{\rm vir}$ \citep{2004A&A...414..445M,2000ApJ...540..113B,2005MNRAS.356.1327G,2009ApJ...692..931L}, we modified the merger tree code to create these links at distances of up to $2.5\,r_{\rm vir}$ (rather than the default $1.0\,r_{\rm vir}$).

\subsection{Coordinates}\label{subsec-coordinates}

We distinguish between two sets of cluster-centric coordinates: $(r,v)$ the full `6D' phase space coordinates, and $(R,V)$ the projected coordinates consisting of the line of sight component of the velocity and the distance to the centre perpendicular to the line of sight. We arbitrarily adopt the third ($z-$)axis of the simulation as our projection axis. The radial projection between two points labelled $1$ and $2$ can be expressed as:
\begin{align}
R_{12} = \sqrt{(r_{2,x}-r_{1,x})^2+(r_{2,y}-r_{1,y})^2}
\end{align}
The projected velocity includes a correction for the Hubble flow, allowing projected coordinates from the simulation to be directly compared to observed line of sight velocity offsets.
\begin{align}
V_{12} = \left|(v_{2,z}-v_{1,z}) + H(r_{2,z}-r_{1,z})\right|
\end{align}
The absolute value encodes our assumption that observationally the distances of clusters and their satellites are not measured with sufficient accuracy to determine the sign of their relative velocity.

To facilitate comparison between clusters, both observed and simulated, we normalize all radial coordinates by the virial radius of the cluster $r_{\rm vir}$, defined using the formula of \citet{1998ApJ...495...80B}: the region enclosing an overdensity 360 times the background density at $z=0$. For those accustomed to a definition in terms of the critical density, an approximate conversion valid at $z=0$ is $r_{200{\rm c}}/r_{\rm vir}\sim0.73$. We normalize velocity coordinates by the 3D velocity dispersion of the cluster, $\sigma_{3{\rm D}}$. We assume that clusters are approximately spherically symmetric so that the observable 1D velocity dispersion $\sigma_{1{\rm D}}$ is $\sigma_{3{\rm D}}\sim\sqrt{3}\sigma_{1{\rm D}}$.

\subsection{Observational sample of clusters}\label{subsec-observed}

\begin{figure*}
\leavevmode \epsfxsize=\columnwidth \epsfbox{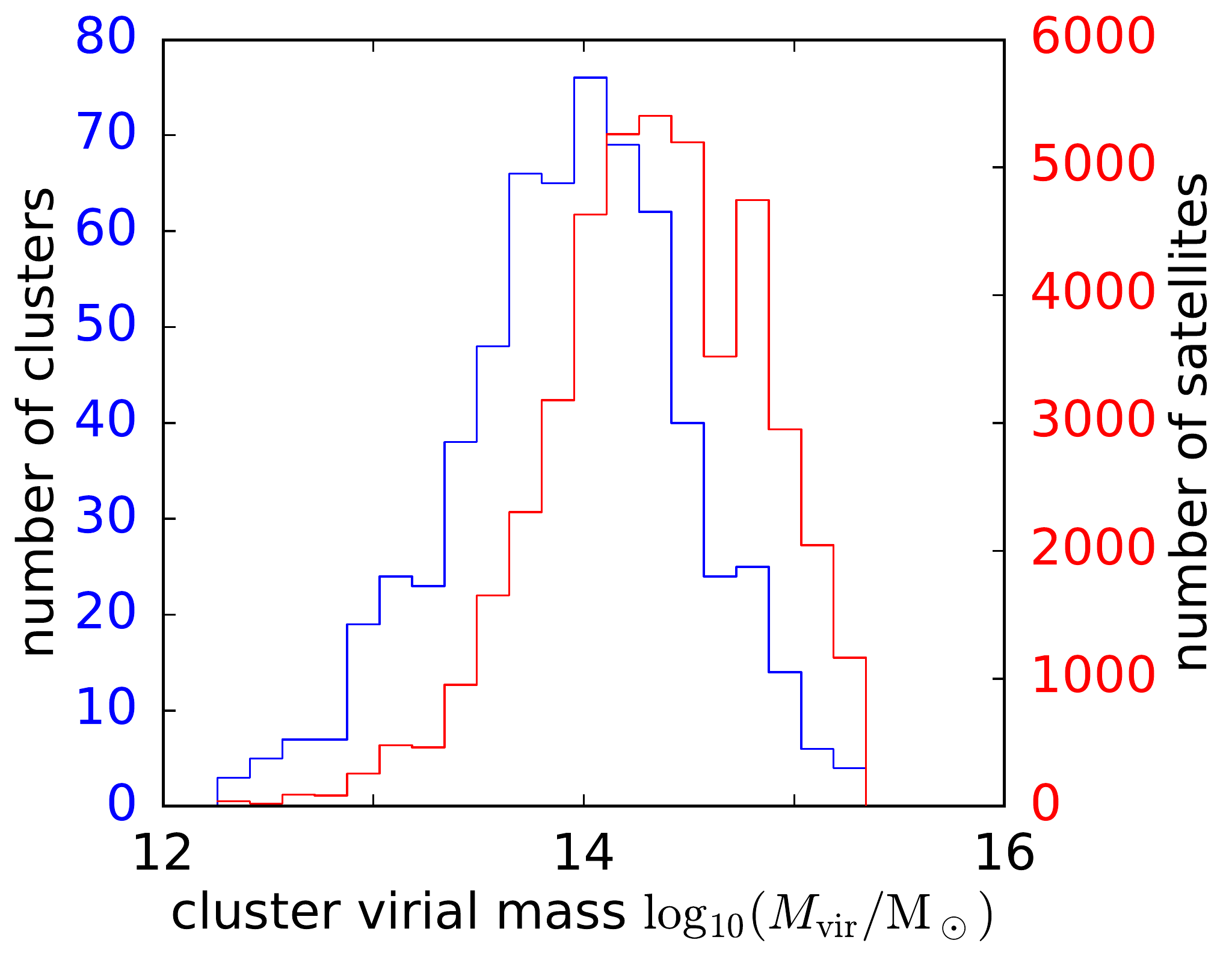}
\hfill
\leavevmode \epsfxsize=.9\columnwidth \epsfbox{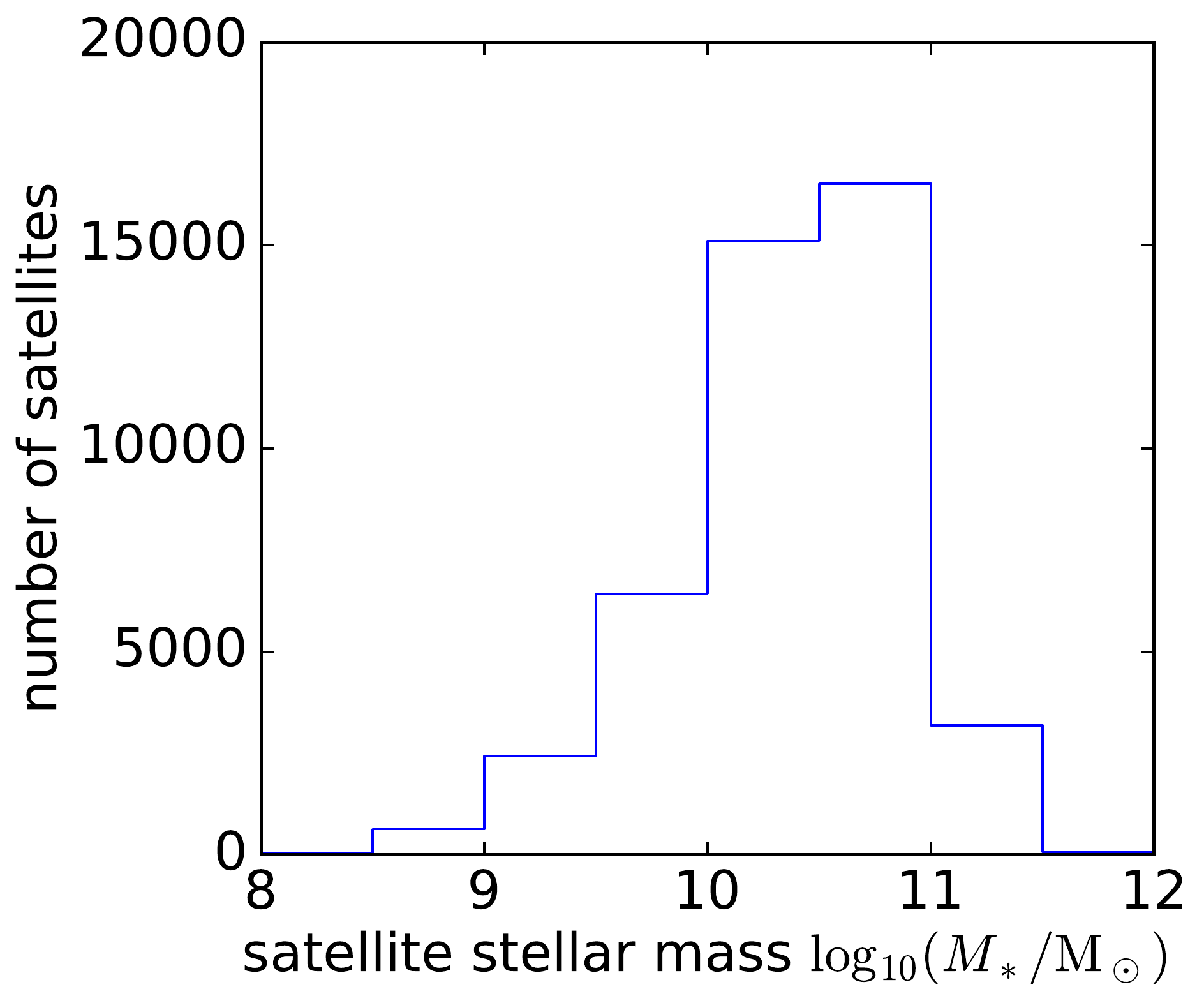}
\caption{\emph{Left panel:} Distribution of host cluster virial masses (dotted blue line) of the observed sample, inferred from the cluster velocity dispersion. Our numerical orbit sample is for hosts of $10^{13} < M_{\rm vir}/\Msun < 10^{15}$. The two datasets are well matched, given that the orbit sample is only very weakly sensitive to host halo mass \citep[see][and Figs.~\ref{mstar_trends},~\ref{fig_compare} and related discussion below]{2013MNRAS.431.2307O}. Also shown is the distribution of satellite galaxies as a function of the cluster mass they occupy (solid red line). \emph{Right panel:} Distribution of satellite candidate stellar masses in the observed sample. Our numerical orbit sample cuts satellite haloes of less than $10^{11.9} \Msun$. An estimate of the halo masses of the satellite candidates in the observed sample using the method of \citet{2014MNRAS.437.2111V} indicates a good match with this mass cut.\label{obs_masses}}
\end{figure*}

To obtain a large sample of clusters and their satellites we use the cluster catalogue of \citet{2007MNRAS.379..867V}. This provides the right ascension, declination, redshift ($z$) and velocity dispersion ($\sigma_{1{\rm D}}$) of 625 clusters. In Fig.~\ref{obs_masses} we show the cluster halo mass distribution of our sample, estimated from a relationship between $\sigma_{1{\rm D}}$ and halo mass calibrated with our simulation sample (\S\ref{subsec-simulations}):
\begin{align}
  \sigma_{\rm 1D}/9.9\times10^{-3}\,{\rm km}\,{\rm s}^{-1}=\left(M_{\rm vir}/\Msun\right)^{0.33}
\end{align}

Our sample of satellites is drawn from the SDSS DR7 \citep{2009ApJS..182..543A}, supplemented with star formation rates \citep[SFR,][]{2004MNRAS.351.1151B,2007ApJS..173..267S} and stellar masses \citep{2014ApJS..210....3M}. We select from the galaxies in this catalogue those with spectroscopic measurements -- and thus more reliable redshifts -- of which there are 562 076.

We construct our sample of observed satellites as similarly as possible to our sample of simulated satellites. The virial radius of each cluster is estimated from its virial mass:
\begin{align}
M_{\rm vir} = \frac{4}{3}\pi r^3_{\rm vir}(360\Omega_m\rho_{\rm crit})
\end{align}
A galaxy is flagged as a satellite candidate of a cluster if it is within $2.5\, r_{\rm vir}$ of the cluster centre and its LoS velocity offset $|\Delta v_{\rm LoS}|$ from the cluster is less than $2.0\,\sigma_{3{\rm D}}$. The brightest cluster galaxies (BCGs) are included in the satellite population in our analysis. We note that by construction, the BCGs have coordinates $(R,V) = (0,0)$ in their respective clusters since \citet{2007MNRAS.379..867V} define the cluster centres in their catalogue as the location of the BCG. This assumes that the BCG is hosted by the cluster halo rather than a satellite halo, and no corresponding orbits would appear in our simulated orbit catalogues. The BCGs account for less than $2$~per~cent of our satellite sample and so we do not expect their presence in the observed sample to impact our conclusions.

The projected radius of the satellites is determined from their angular separation from their host cluster centre $\Delta\theta$:
\begin{align}
\frac{R}{r_{\rm vir}} = \frac{d_A\Delta\theta}{r_{\rm vir}}
\end{align}
$d_A$ is the angular diameter distance of the cluster. The velocity offset is calculated from the redshift offset:
\begin{align}
\frac{V}{\sigma_{3{\rm D}}} = \frac{c|z_g-z_c|}{(1+z_c)\sqrt{3}\sigma_{1{\rm D}}}
\end{align}
$z_g$ is the redshift of the galaxy and $z_c$ is the redshift of the cluster.

This process of associating galaxies to clusters yields a sample of 44 436 satellite candidates, of which we expect about half to be interlopers (see \S\ref{subsec-interlopers}). The distribution of satellite stellar masses is shown in Fig.~\ref{obs_masses}. We estimate the halo mass of the satellites, based on the method of \citet{2014MNRAS.437.2111V}, and the conversion from $M_{200{\rm c}}$ to $M_{\rm vir}$ of \citet{2001MNRAS.321..559B}, as:
\begin{align}
\log_{10}\left(M_{\rm vir}/\Msun\right) = 0.84\log_{10}\left(M_{*}/\Msun\right)+3.09
\end{align}
We estimate a systematic error in this conversion of up to $15$~per~cent, and a scatter of up to $0.4\,{\rm dex}$. 

\section{Method}\label{sec-method}

\subsection{Orbit Libraries}\label{subsec-orbits}

We use the same method to construct orbit libraries as \citet{2013MNRAS.431.2307O}, which we summarize here. We define clusters as haloes of $>10^{13} \Msun$ at $z=0$, of which MDR1 has $\sim 550\,000$. Satellites are within $2.5\,r_{\rm vir}$ of a cluster at $z=0$, and have a mass $>10^{11.9} \Msun$ \emph{at infall} such that they are well-resolved and minimally sensitive to artificial disruption \citep{1999ApJ...522...82K,2008MNRAS.391.1489K}. We track the satellites back in time, noting the time of infall which we define as entry of the satellite within $2.5\,r_{\rm vir}$. This somewhat unconventional definition of infall has the advantage that satellites nearly never orbit back out past this radius. A satellite on a typical orbit takes $2.5-3.0$~Gyr after infall to reach the virial radius, and pericentre occurs $3.5-4.0$~Gyr after infall. Our final sample numbers $\sim1\,200\,000$ satellite orbits.

\subsection{Interlopers}\label{subsec-interlopers}

\begin{figure}
\leavevmode \epsfxsize=\columnwidth \epsfbox{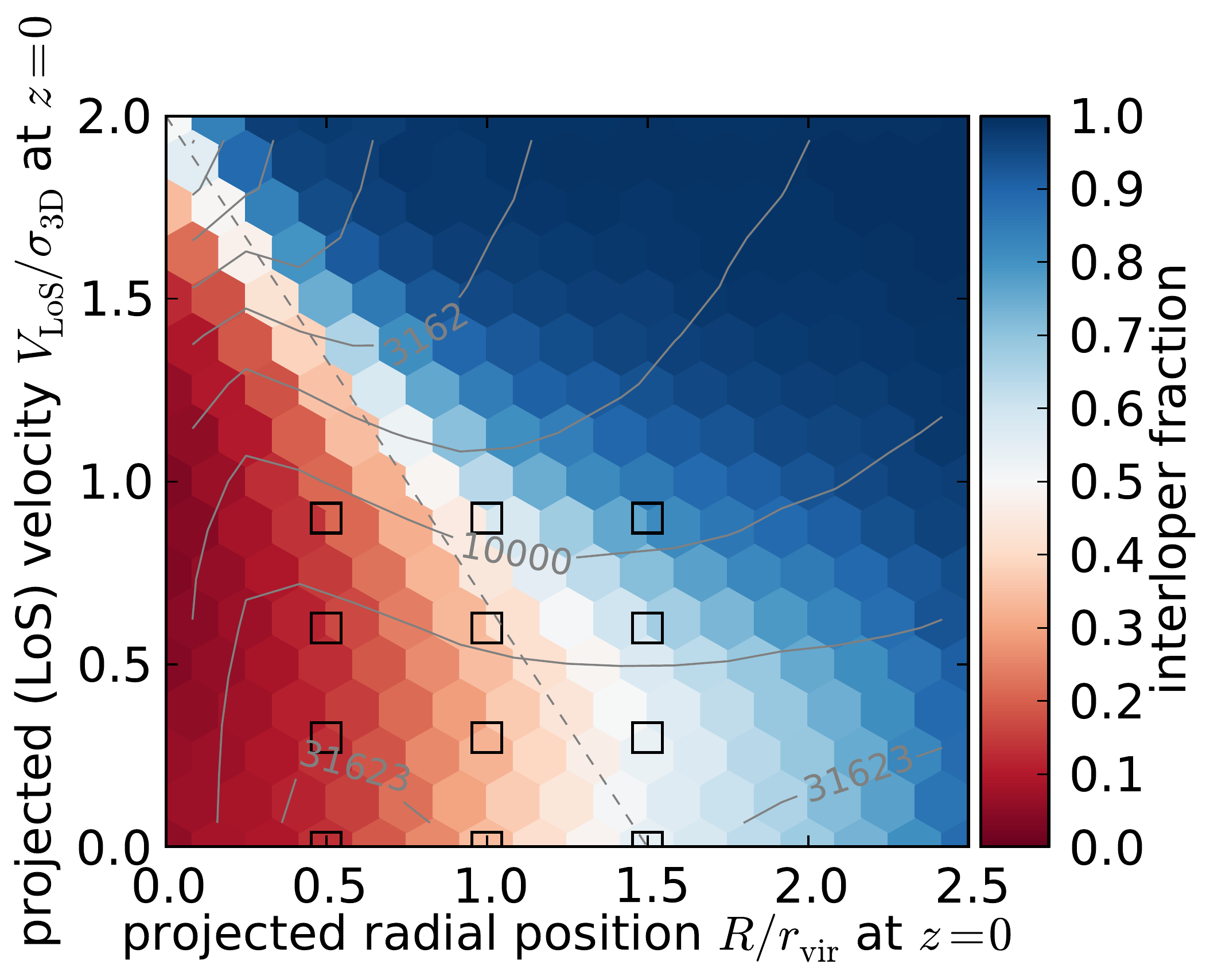}
\caption{Fraction of `interlopers' -- satellite haloes which appear in a cluster in projection with $R<2.5\,r_{\rm vir}$ and $V<2.0\,\sigma_{3{\rm D}}$, but fall outside in 3D, that is $r > 2.5\,r_{\rm vir}$ -- as a function of position in PPS. The halo population at small projected radii and velocities is dominated by bona-fide cluster members, while at large projected radii and velocities the interloper fraction tends to 1.0. The dashed line marks $\frac{V}{\sigma_{3{\rm D}}}=-\frac{4}{3}\frac{R}{r_{\rm vir}}+2$ and approximately divides the two regions. Contours indicate the number of haloes in each bin. The black squares correspond to the locations of the subsamples shown in the panels in Fig.~\ref{sample_pdfs}.\label{misidentification_fraction}}
\end{figure}

An observed sample of cluster satellites is typically selected within some projected radius and velocity offset from the cluster centre. This defines a cylinder\footnote{More accurately the shape is that of a cone with its peak sliced off, but a cylinder is a good approximation for distant clusters.} in PPS. For reasonable selection cuts, this cylinder encloses a sphere centered around the cluster which contains the satellite galaxies of interest, but also a volume outside this sphere containing galaxies `projected into' the cluster, which we term `interlopers'. Some of these interlopers will likely someday fall into the cluster, while others may eventually move off into a neighbouring structure. We supplement our simulated orbit catalogue with a sample of interlopers, allowing a fair comparison with our sample of observed cluster satellite candidates. We select all haloes with projected coordinates (projection along the $z-$axis of the simulation box) $\frac{R}{r_{\rm vir}} < 2.5$, $\frac{V}{\sigma_{3{\rm D}}} < 2.0$ that also have non-projected radius $\frac{r}{r_{\rm vir}} > 2.5$. We apply the same mass cut as for our simulated satellite sample, yielding $\sim1\,500\,000$ interlopers. The fraction of interlopers (compared to actual satellites) is a function of position in PPS, with interloper fraction increasing with increasing $R$ and $V$ (see Fig.~\ref{misidentification_fraction}).

\begin{figure*}
\leavevmode \epsfxsize=2\columnwidth \epsfbox{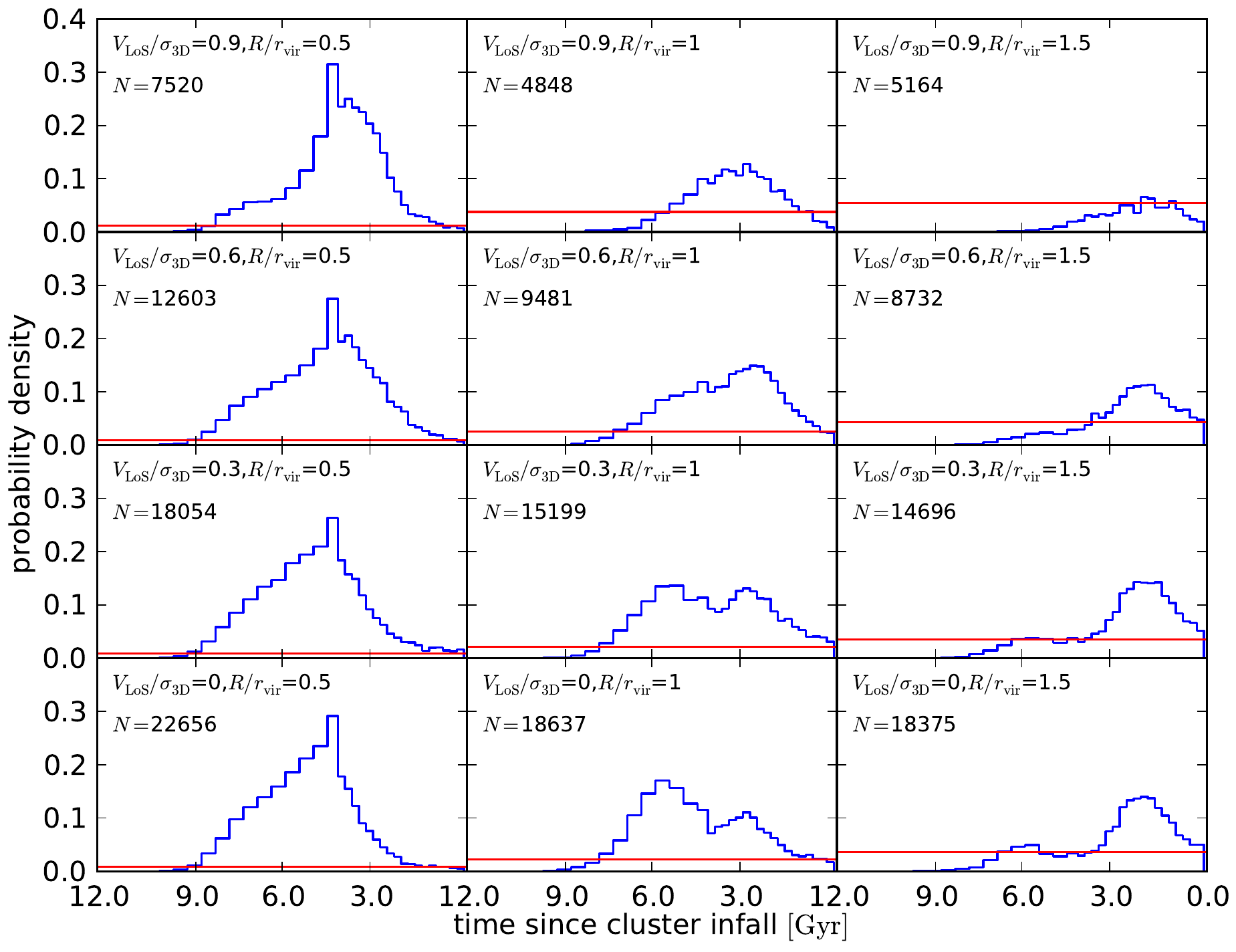}
\caption{Probability density functions (PDFs) of time since cluster infall for a selection of points in PPS. The panels correspond to the locations marked by black squares in Fig.~\ref{misidentification_fraction}. The red line corresponds to the interloper probability: the ratio of the integrals of the red and blue curves is equal to the interloper fraction. The number of orbits used to construct each histogram is labelled $N$, giving a measure of the statistics of each PDF.\label{sample_pdfs}}
\end{figure*}

From our catalogue of satellite orbits and interlopers, we construct a probability density of infall times for each position in PPS. Example probability density functions for a selection of points in the $(R,V)$ plane (see boxes in Fig.~\ref{misidentification_fraction}) are shown in Fig.~\ref{sample_pdfs}.

\subsection{Comparison of observed and simulated samples}

\begin{figure}
\leavevmode \epsfxsize=\columnwidth \epsfbox{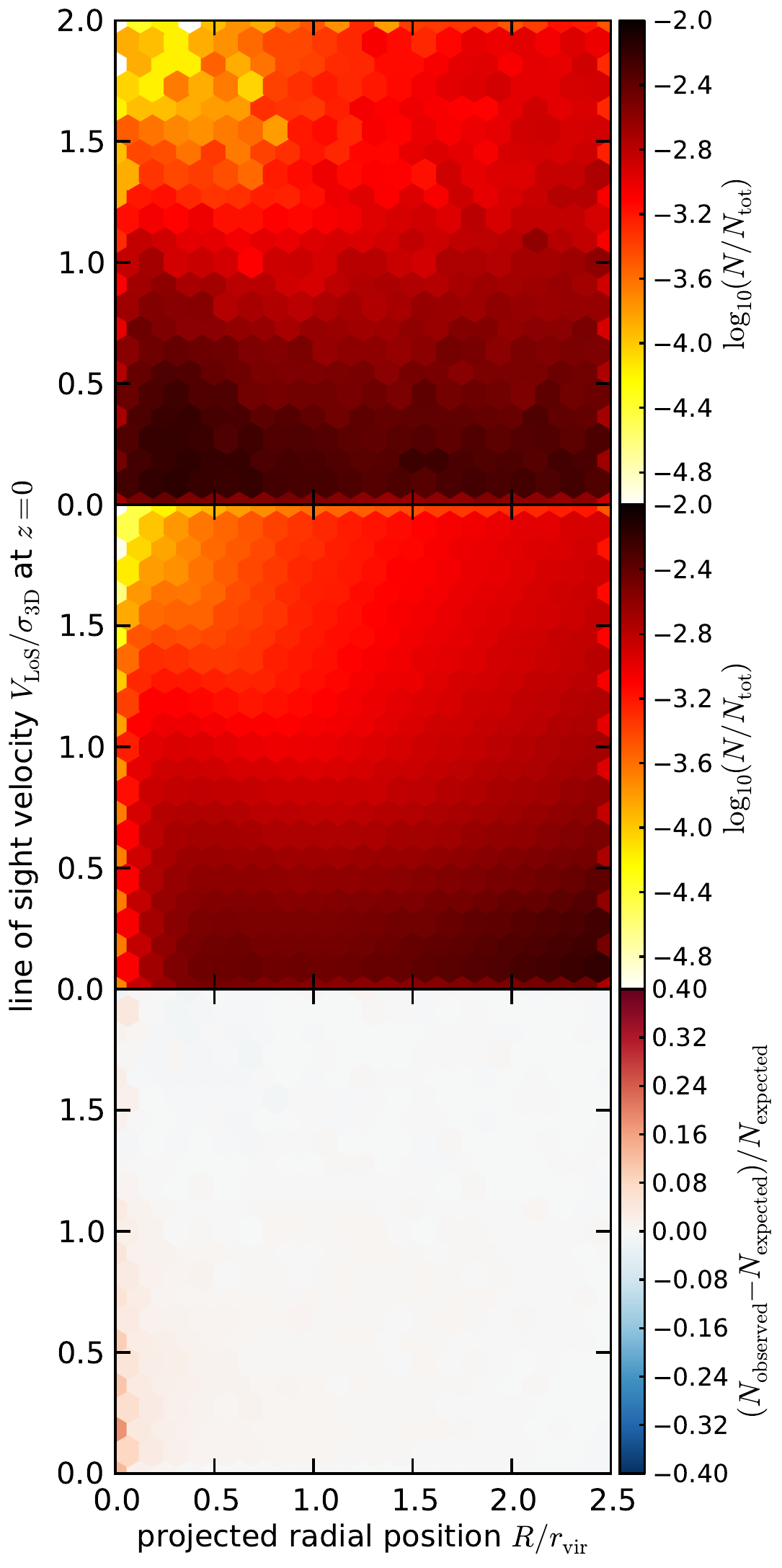}
\caption{\emph{Upper panel:} Distribution of observed galaxy sample in PPS. \emph{Middle panel:} Distribution of simulated halo sample in PPS. The nomalization the two upper panels is such that the colour scales are directly comparable. \emph{Lower panel:} Fractional excess of observed galaxies relative to expectation from simulation halo counts as a function of PPS position. Most regions are limited to variations of a few per~cent. The excess of observed galaxies at low $R$ and $V$ is consistent with the amount of artificial disruption of simulated haloes near the centres of clusters \citep{2008MNRAS.391.1489K,1999ApJ...522...82K}; a smaller proportion of observed galaxies are also absent due to fiber collisions in the SDSS.\label{match_sample}}
\end{figure}

The observed and simulated catalogues of satellite candidates (i.e. including interlopers in both cases) are generally well matched. The distribution of observed cluster virial masses (Fig.~\ref{obs_masses}) is similar to our cutoff of $>10^{13} \Msun$ in the simulations, and the distribution of infall times as a function of $(R,V)$ coordinates is only weakly sensitive to host mass \citep[][and Figs.~\ref{mstar_trends},~\ref{fig_compare} and related discussion below]{2013MNRAS.431.2307O}. The satellite halo mass distributions are also well matched, with the offset between the observed and simulated samples being comparable to the estimated systematic error in the conversion from stellar mass to halo mass. Finally, the PPS distributions of the simulated and observed haloes are in excellent agreement, as shown in Fig.~\ref{match_sample}.

\subsection{Quenching model}\label{subsec-models}

In this section we briefly derive the likelihood function of our simple quenching model. We assume that when quenching occurs, individual galaxies transition rapidly from an active, star forming state to a passive, quenched state. This is motivated by the bimodal distribution of SSFR (see for instance Fig.~\ref{ssfr_cut}), also apparent as the `green valley' in the colour distribution of galaxies \citep{2003MNRAS.341...33K, 2004ApJ...600..681B}. We assume that two galaxy populations exist, one inside and the other outside (but nearby) massive clusters, each with distinct passive fractions. We fit two timescales: the first is a delay between cluster infall and the onset of quenching, the second is the timescale for the transition between the quenched fraction of the population just outside the cluster to that inside.

The likelihood $\mathcal{L}$ is defined in terms of a sum over probabilities $P_i$ where $i$ is an index running over all galaxies in the observed sample. Each galaxy has three properties of interest in the context of calculating the likelihood: a projected radius from its host cluster $R_i$, a line of sight velocity offset from the host cluster velocity $V_i$ and a specific star formation rate ${\rm SSFR}_i$.
\begin{align}
\ln \mathcal{L} = \sum_i \ln P_i
\end{align}
If a galaxy is observed to be passive, with ${\rm SSFR}_i < {\rm SSFR}_{\rm cut}(M_*)$ (we adopt a cut in SSFR between active and passive galaxies that depends on stellar mass, see \S~\ref{sec-def-passive} and Fig.~\ref{sfms}), then $P_i$ is the probability according to the model that the galaxy is passive $p_{\rm passive}$. Conversely, if the observed galaxy is active, $P_i$ is the probability according to the model that the galaxy is active, $(1-p_{\rm passive})$.
\begin{align}
P_i = 
\begin{cases}
p_{{\rm passive},i}, \: & {\rm SSFR}_i \leq {\rm SSFR}_{\rm cut}(M_*) \\
1-p_{{\rm passive},i}, \: & {\rm SSFR}_i > {\rm SSFR}_{\rm cut}(M_*)\\
\end{cases}
\end{align}
$p_{{\rm passive},i}$ is in turn defined in terms of the passive fraction outside the cluster $f_{\rm passive,out}$ and inside the cluster $f_{\rm passive,in}$, which are related by $\Delta f_{\rm passive} = f_{\rm passive,in} - f_{\rm passive,out}$, and the probability $p_{q,i}$ that the cluster has quenched the galaxy.
\begin{align}
p_{{\rm passive},i} = f_{\rm passive,out} + p_{q,i} \Delta f_{\rm passive}
\end{align}
$p_{q,i}$ is expressed as the integral of the product of two probabilities. The first is the probability $p_q(t)$ that at time $t$ after infall the cluster has quenched the satellite. The functional form of $p_q(t)$ is an input of the model. The second is the probability $p_{{\rm infall},i}(R_i,V_i,t)$ that the galaxy has a time since infall of $t$.
\begin{align}
p_{q,i} = \int_{t=0}^{t_f} p_q(t)p_{{\rm infall},i}(R_i,V_i,t){\rm d}t
\end{align}
$t_f$ is the age of the universe. $p_{{\rm infall},i}$ is extracted from our orbit libraries, as illustrated in Fig.~\ref{sample_pdfs}.

We choose $p_q(t)$ to represent a scenario where a galaxy is unaffected after infall until a time $\Delta t$ has elapsed, then has an increasing probability of being quenched, parameterized by a timescale $\tau$. The interpretation of $\tau$ is as a scatter in the quenching time of a \emph{population} of satellites, rather than the time taken for an individual galaxy to transition from active to passive. We assume that this transition occurs rapidly, motivated by the bimodality in the ${\rm SSFR}$ distribution \citep[Fig.~\ref{sfms}, see also][]{2012MNRAS.424..232W}, and so make no attempt to model the full ${\rm SSFR}$ distribution at each point in PPS, instead reducing this distribution to a single parameter, the passive fraction.
\begin{align}
p_q(t) =
\begin{cases}
0, \: & t \leq \Delta t\\
1-e^{-(t-\Delta t)/\tau} \: & t > \Delta t
\end{cases}
\end{align}
The model described above has $4$ free parameters to be fit: $\Delta t$, $\tau$, $f_{\rm passive, out}$ and $\Delta f_{\rm passive}$. When discussing our results below, we often quote $f_{\rm passive, in}$ instead $\Delta f_{\rm passive}$, which is equivalent, but, we feel, more intuitive. We also consider a single `combined' timescale $t_{1/2}=\Delta t + 0.69\tau$, the time when half of the galaxies that will be quenched by their host have become passive. In order to derive formal errors on this parameter, we perform additional fits where this parameter is fit directly (replacing $\Delta t$ as a parameter to be fit).

We adopt flat priors on all parameters in the intervals $0 \leq f_{\rm passive, in} \leq 1$, $0 \leq f_{\rm passive,out} \leq 1$, $\Delta t \geq 0$~Gyr, $\tau \geq 0$~Gyr. In cases where we fit $t_{1/2}$ instead of $\Delta t$, we constrain $t_{1/2} > 0$~Gyr.

\subsection{Definition of `active' and `passive'}\label{sec-def-passive}

We split our sample of cluster satellite candidates into an active and a passive population based on their SSFRs. The distribution of SSFRs shows a clear bimodality, and the relative size of the two populations is a function of PPS coordinates (see the upper left panel of Fig.~\ref{model_composite}). The relative size of the passive population increases with decreasing $R$ and $V$.

\begin{figure}
\leavevmode \epsfxsize=\columnwidth \epsfbox{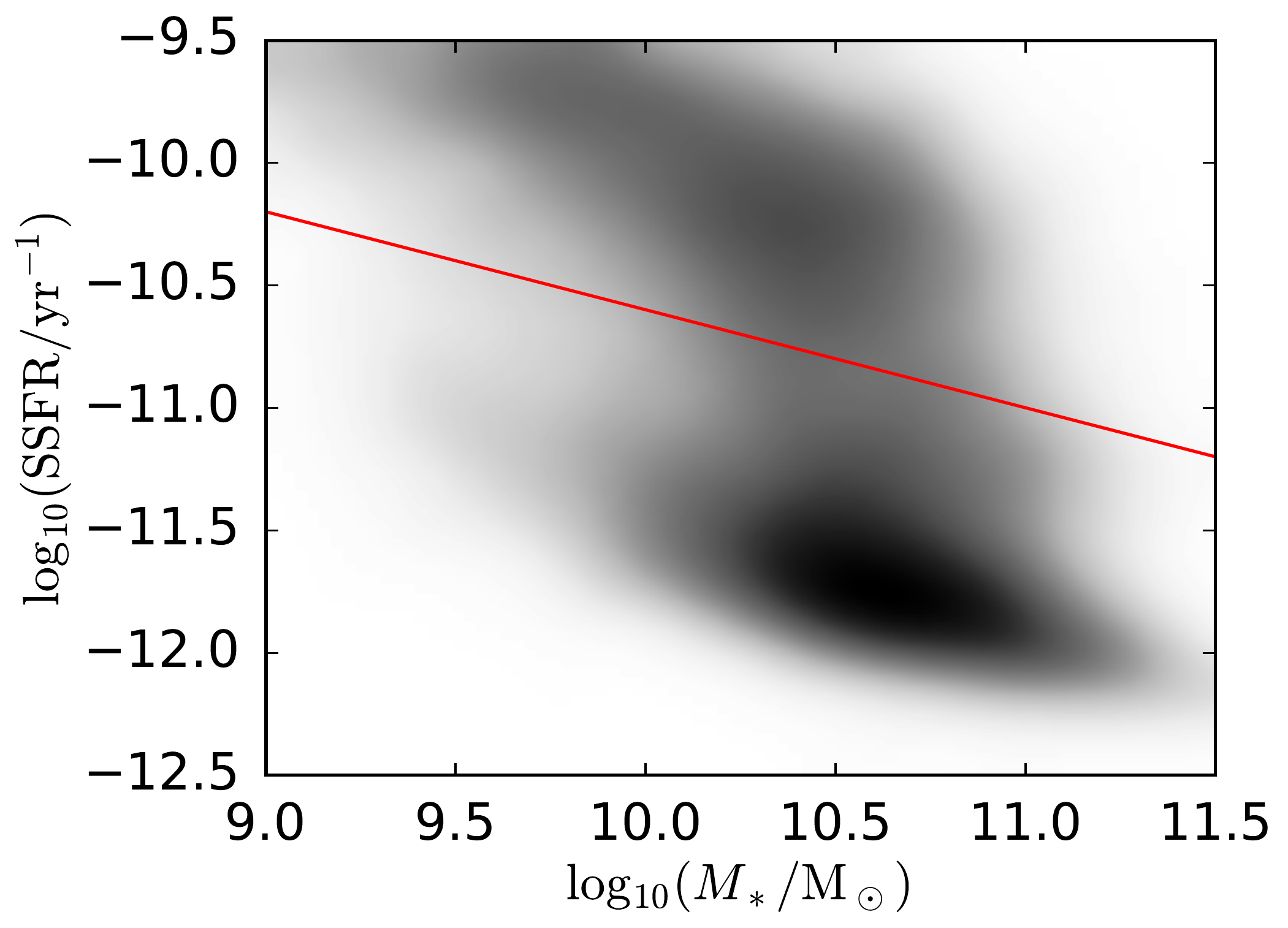}
\caption{SSFR as a function of stellar mass for our sample of galaxies. The colour scale is logarithmic and represents the density of galaxies in this plane. The passive `red sequence' and active `blue cloud' populations are clearly visible. We separate the two `by eye' with a line (see equation~\ref{ssfr_cut}) to define our ${\rm SSFR}_{\rm cut}(M_*)$. \label{sfms}}
\end{figure}

Whether a galaxy with a given SSFR should be classified as active or passive depends on its stellar mass $M_*$. Fig.~\ref{sfms} shows the distribution of SSFRs as a function of $M_*$ for our sample of galaxies. The separation between the active and passive populations was determined `by eye' and is illustrated by the red line at:
\begin{align}
\log_{10}({\rm SSFR}_{\rm cut}/{\rm yr}^{-1}) &= -\frac{\log_{10}(M_*/\Msun)}{2.5} - 6.6 \label{ssfr_cut}
\end{align}

\section{Results}\label{sec-results}

In this section we fit our simple quenching model to the observed sample of satellite galaxies. We split the observed data sample into two bins of host halo mass, $10^{13}-10^{14}$ and $10^{14}-10^{15} \Msun$, and five bins of satellite stellar mass, evenly spaced in $\log_{10}(M_*/\Msun)$ between $9$ and $11.5$. For each of the ten subsamples we produce an infall time PDF from our orbit libraries (including interlopers) using only orbits consistent with the host and satellite mass corresponding to each bin.

\begin{figure*}
\leavevmode \epsfxsize=2\columnwidth \epsfbox{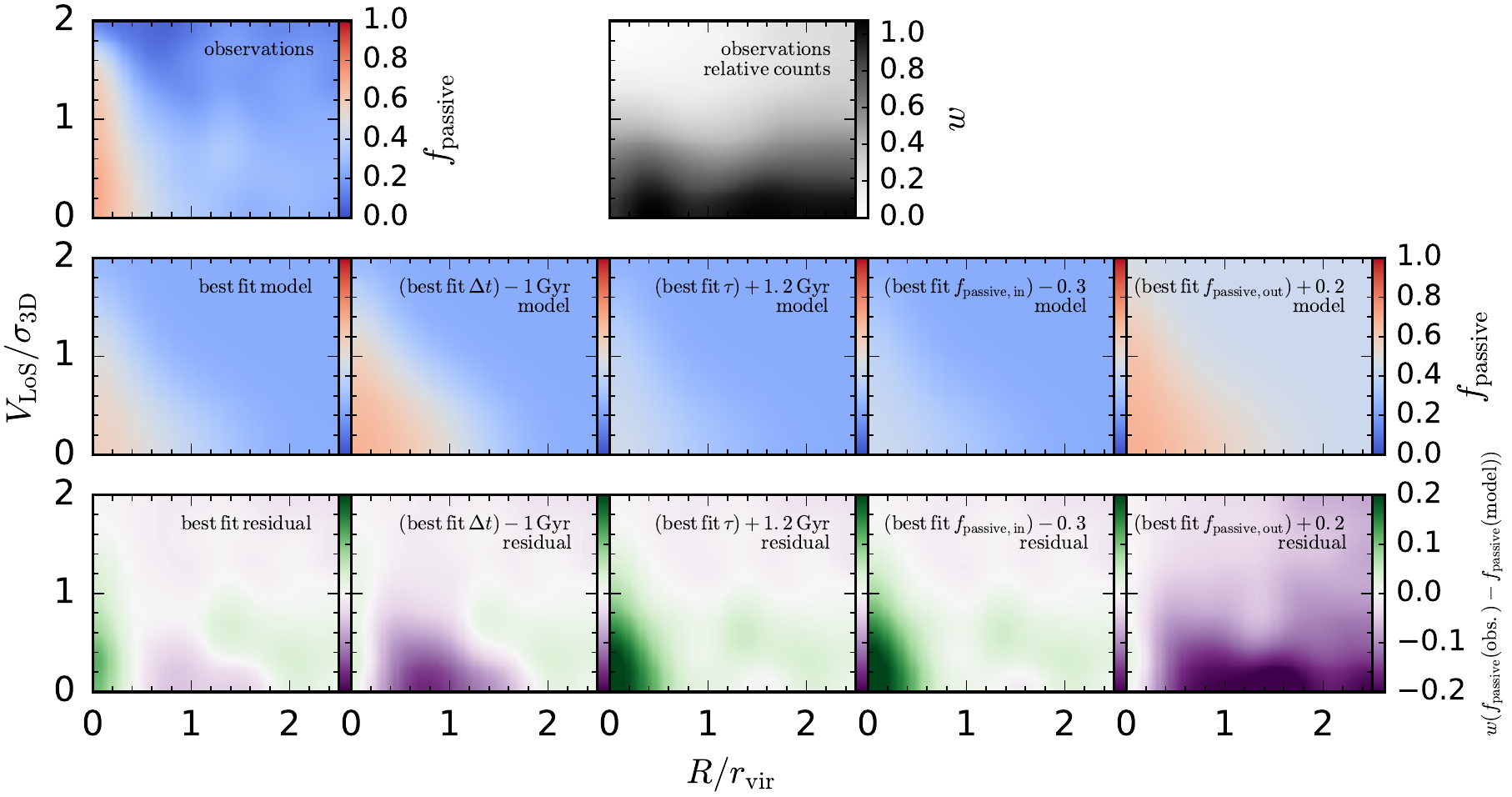}
\caption{A visual representation of our quenching model. The top left panel shows the observed passive fraction as a function of phase space coordinates for galaxies in the ranges $14<\log_{10}(M_{\rm host}/\Msun)<15$ and $9.5<\log_{10}(M_*/\Msun)<10$, smoothed using a gaussian kernel with $\sigma_V=\sigma_R=0.2$ (all other panels use the same smoothing). The top right panel shows the smoothed relative counts of observations in the phase space plane, which we denote $w$. The middle row shows the prediction of our model the parameter choices yielding the maximum likelihood (leftmost panel), and the effect of changing each of the parameters on the predictions; from left to right: $\Delta t$, $\tau$, $f_{\rm passive, in}$, $f_{\rm passive, out}$. The bottom row shows the difference between the observations and the model predictions for each set of parameters from the second row, weighted by the relative counts: $w(f_{\rm passive}({\rm observations})-f_{\rm passive}({\rm model}))$.\label{model_composite}}
\end{figure*}

A visual depiction of our model (see \S\ref{subsec-models}), illustrated using the $14<\log_{10}(M_{\rm host}/\Msun)<15$ and $9.5<\log_{10}(M_*/\Msun)<10$ subsample, is presented in Fig.~\ref{model_composite}. Given the four parameters $f_{\rm passive,out}$, $f_{\rm passive,in}$, $\Delta t$ and $\tau$, the model predicts the observable $f_{\rm passive}$ as a function of position in PPS. In the middle row of panels in Fig.~\ref{model_composite}, the leftmost panel shows the prediction of the model for our best fit parameter values. The next four panels show the effect on the prediction of changing individual parameters. The observed distribution of $f_{\rm passive}$ in the $(R,V)$ plane is shown in the upper left panel, and the density of galaxies $w$ as a function of position in the $(R,V)$ plane, relative to the maximum density, is shown in the upper right panel. The likelihood of a given set of parameter values reflects the comparison between the model prediction and the observed data. The third row shows a visualization of this comparison: darker colour is used where there is a larger discrepancy between model and data, weighted by $w$.

\begin{figure*}
\leavevmode \epsfxsize=2\columnwidth \epsfbox{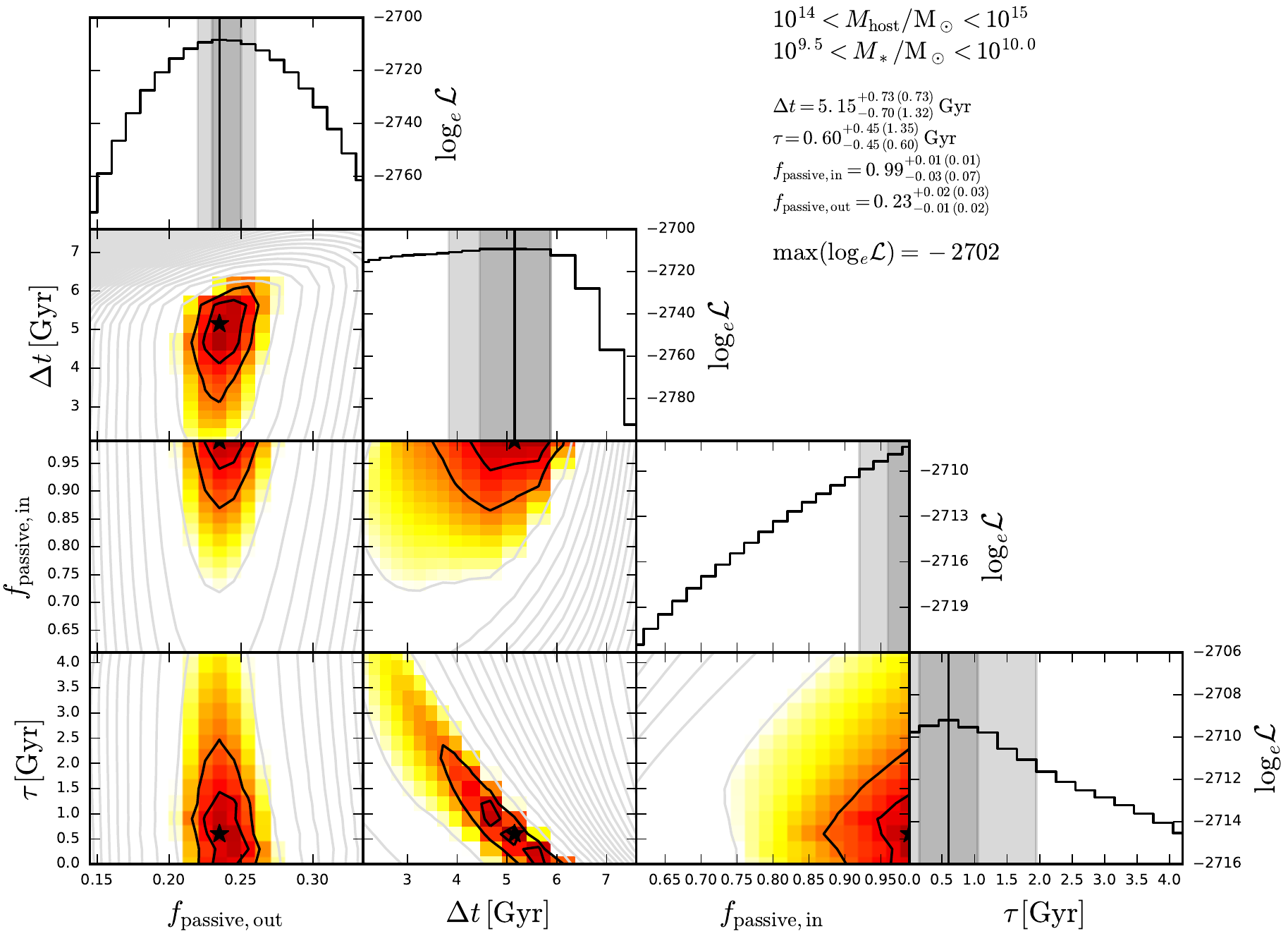}
\caption{Marginalized likelihood distributions for individual model parameters (panels along the diagonal) and marginalized likelihood maps for pairs of model parameters (off-diagonal panels), for the data in the ranges $10^{14}<\log_{10}(M_{\rm host}/\Msun)<10^{15}$ and $10^{9.5}<\log_{10}(M_{\rm *}/\Msun)<10^{10.0}$. The location of the global maximum likelihood (i.e. without marginalization) is shown with a vertical black line (histograms) or a black star (maps). The $68$ and $95$~per~cent confidence intervals are indicated by the dark and light shaded regions (histograms) and the inner and outer black contours (maps), respectively. In the maps, the pale gray contours show the overall shape of the likelihood distribution, with each contour representing an additional drop of $3$~per~cent relative to the maximum. The detailed distribution around the maximum is instead shown with a colour scale. The intervals given in the text labels are $68$ ($95$)~per~cent confidence intervals.\label{banana}}
\end{figure*}

The time resolution of the MDR1 simulation snapshots imposes a limit on the precision with which we can measure $\Delta t$, so we only evaluate $\mathcal{L}$ for values of $\Delta t$ spaced to match the simulation time resolution. The discrete nature of $\Delta t$ makes a simultaneous multidimensional maximum likelihood search (e.g. Markov Chain Monte Carlo) somewhat unwieldly. Instead, we use an iterative grid search strategy, beginning with a coarse $4$-dimensional grid covering a wide area of our parameter space, and refining until we adequately sample the peak and $95$~per~cent confidence region of the likelihood distribution. The result of this fitting for the same subsample used illustratively in Fig.~\ref{model_composite} is shown in Fig.~\ref{banana}. The panels along the diagonal show the posterior distributions for each of the four parameters, marginalized over all other parameters in each case. The dark (light) gray shaded regions show the $68$~per~cent ($95$~per~cent) confidence intervals, and the solid vertical line indicates the parameter value at the global likelihood maximum (which may be distinct from the maximum of the marginalized distribution for a single parameter). The off-diagonal panels show the marginalized posterior distributions for pairs of model parameters. The two black contours show the $68$ and $95$~per~cent confidence intervals, and the star symbol corresponds to the location of the global maximum likelihood (again, possibly distinct from the maximum of any given maginalized distribution). The other nine fits give qualitatively similar results; one key recurring feature of the posterior distributions is the degeneracy between $\Delta t$ and $\tau$, with shorter quenching times $\Delta t$ being `compensated' by longer transition timescales $\tau$, which will be discussed further in \S\ref{sec-conclusions}.

\begin{figure*}
\leavevmode \epsfxsize=2\columnwidth \epsfbox{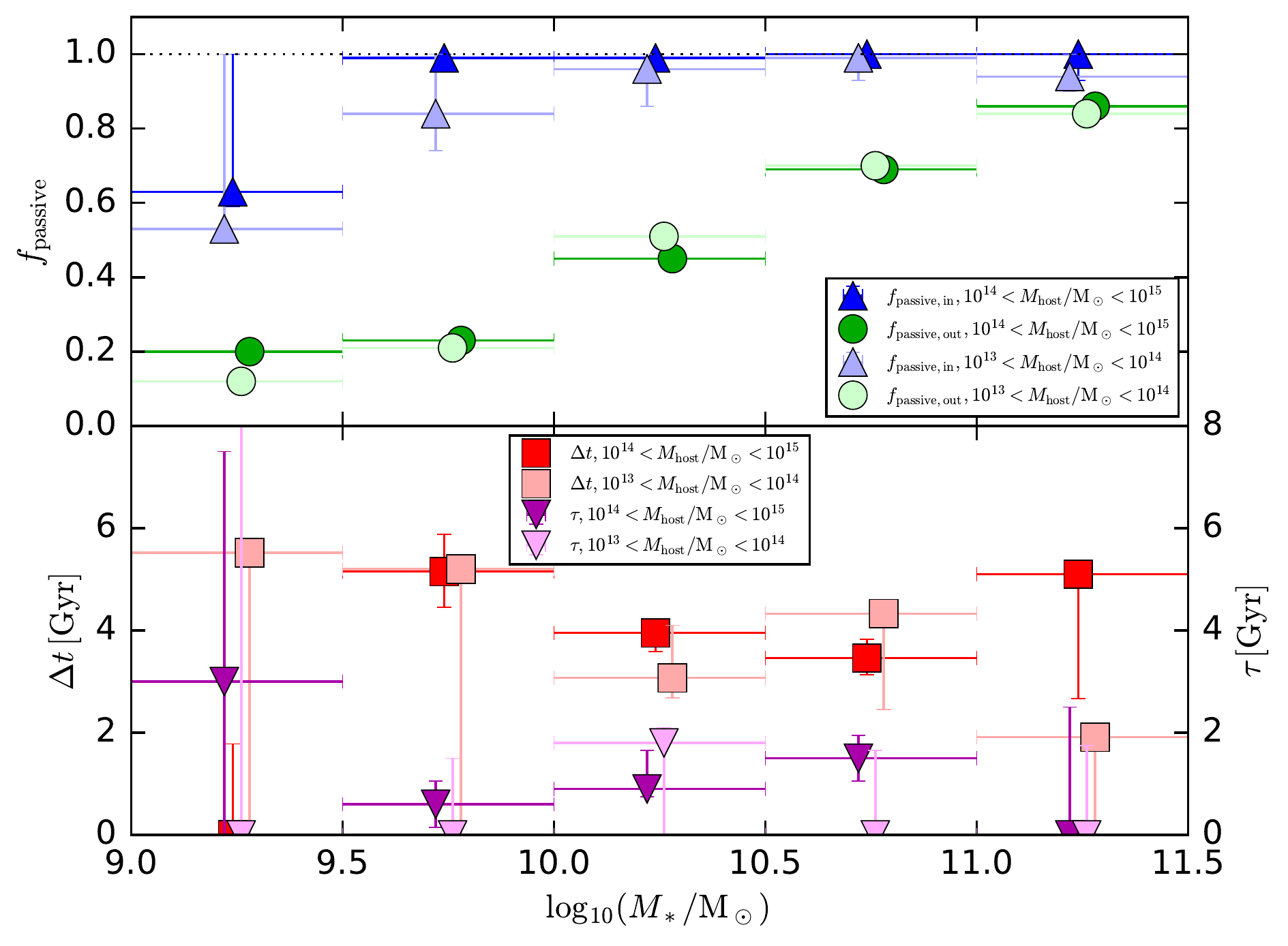}
\caption{Maximum likelihood model parameters (upright triangles: $f_{\rm passive,in}$; circles: $f_{\rm passive, out}$; squares: $\Delta t$; inverted triangles: $\tau$) for the observational sample binned by stellar mass for host masses in the range $10^{13}-10^{14} \Msun$ (pale symbols) and $10^{14}-10^{15} \Msun$ (dark symbols). The horizontal error bars show the bin widths, the vertical error bars the $68$~per~cent confidence intervals. For clarity, the symbols have been slightly offset horizontally.\label{mstar_trends}}
\end{figure*}

In Fig.~\ref{mstar_trends} and Table~\ref{results_table} we summarize the results of fitting our model to each of the ten data subsamples. All vertical error bars and quoted intervals represent $68$~per~cent confidence intervals (which in some cases are smaller than the symbols or run off the region shown); horizontal error bars simply indicate bin widths.

\begin{table*}
  \caption{Summary of our results. The first two columns indicate the host mass and stellar mass bins. The next four columns list the parameter choices yielding the maximum likelihood and the $68$~per~cent confidence intervals for each. The penultimate column lists $t_{1/2}$ (see text), including the $68$~per~cent confidence interval derived by marginalizing the posterior distribution over this parameter combination. Because we perform a new fit to derive errors on this parameter, and because of the approximate nature of a maximum likelihood grid search, the values are not always exactly equal to $\Delta t + 0.69 \tau$, but \emph{are} consistent within our quoted errors. The last column lists the maximum likelihood.\label{results_table}}
  \input{figs/table} 
\end{table*}

\section{Discussion}\label{sec-conclusions}

In the results of fitting our model, shown in Fig.~\ref{mstar_trends}, we note the expected trend in $f_{\rm passive,out}$, with higher stellar mass galaxies outside the clusters that are more affected by `internal quenching' having higher $f_{\rm passive,out}$. Interestingly, in all cases we recover values of $f_{\rm passive, in}$ consistent with $1.0$ within the $68$~per~cent confidence interval, and in most cases the best fit value is $\approx 1.0$, suggesting that quenching by clusters is $100$~per~cent efficient\footnote{Indicating that, once quenching by the cluster has had time to operate, $100$~per~cent of satellites have been quenched, but not that the cluster is responsible for quenching $100$~per~cent of the passive galaxies it contains.}, in agreement with the conclusions of \citet{2011MNRAS.416.2882M}. Of course, the \emph{observed} fraction of quenched satellites will always be less than this, even in radial bins closest to the cluster centre, because these bins contain a fraction of satellites falling into the cluster for the first time, which have not yet had time to be quenched, and galaxies `projected into' the cluster.

We find a flat or perhaps slightly decreasing trend in $\Delta t$ with increasing stellar mass, and usually small values of $\tau$ (however, in a few cases up to several Gyr, though with $95$~per~cent confidence intervals still consistent with near-zero values). We examine the details of these trends in more detail below.

In the upper panel of Fig.~\ref{mstar_trends}, the lower best fit values of $f_{\rm passive,in}$ in the lowest $M_*$ bin stand out as peculiar (also, to a lesser extent, for the lower $M_{\rm host}$ bin in the $9.5<\log_{10}(M_*/\Msun)<10$ bin). While this could be a sign that these lower mass galaxies are more resiliant to quenching, considering other peculiarities in the fits in these bins we cautiously prefer an interpretation where $f_{\rm passive, in}\approx1.0$ in all cases. We first point out that the $68$~per~cent confidence intervals extend up to $1.0$ in all cases. In all cases the marginalized posterior probability distribution for $f_{\rm passive,in}$ (not shown) peaks at $1.0$, but in these peculiar cases the global maximum likelihood is offset from the peak of the marginalized distribution. We believe that this is due to the proximity to the mass resolution limit in the simulations, which causes the satellite haloes contributing orbits for use in this bin to be biased more toward the upper edge of the bin than they would otherwise be. \citet{2013MNRAS.431.2307O} showed that higher mass satellite haloes, which host higher mass galaxies, have orbits with preferentially smaller backsplash distances, which is easily understood as the effect of dynamical friction. If a PDF constructed from a collection of satellite orbits biased toward higher masses is used, when fitting the model, the lower mass galaxies have inferred times since infall that are biased low, driving down the fit quenching timescales ($\Delta t$, $\tau$, or both). This picture seems consistent with the timescales plotted in the lower panel of Fig.~\ref{mstar_trends}, particularly for the higher $M_{\rm host}$ bin, which has a seemingly unrealistic\footnote{Keeping in mind our definition of infall at $2.5\,r_{\rm vir}$} best-fit $\Delta t=0$. This has a knock-on effect on $f_{\rm passive,in}$, driving the best fit to lower values. This situation is exacerbated by the relatively low numbers of observed cluster satellite candidates in these mass bins (see Fig.~\ref{obs_masses}). These difficulties are reflected in the statistical uncertainties derived from the posterior distribution; with the exception of those for $f_{\rm passive,out}$, which is constrained primarily by the properties of interlopers, these are very large.

The trends seen in the $\tau$ parameter are also puzzling at first glance. This parameter turns out to be difficult to constrain using our methodology, with $68$~per~cent confidence intervals up to several Gyr wide. Inspecting the marginalized posterior distributions (e.g. Fig.~\ref{banana}), we invariably find a strong degeneracy between $\tau$ and $\Delta t$. This is intuitive, as a rapid transition at a given time is numerically similar to a slightly slower transition that begins slightly earlier. We obtain tighter constraints by considering a representative single timescale $t_{1/2}$. The trends and $68$~per~cent confidence intervals for this parameter combination for the higher (lower) $M_{\rm host}$ bin as a function of $M_*$ are illustrated by the solid (dashed), red (pink) lines and corresponding shaded regions in Fig.~\ref{fig_compare}. The intervals remain large for the subsamples with relatively low observed galaxy counts, but we verify via fits to a Monte Carlo sampling of the quenching timescale distributions that the decreasing trend with increasing $M_*$ is significant at $92$~per~cent ($61$~per~cent) confidence for the lower (higher) $M_{\rm host}$ bin. Further efforts to understand these trends would likely benefit from a more sophisticated model which explicitly models the trends and fits data across the entire $M_*$ and $M_{\rm host}$ range simultaneously.

We performed two tests to investigate the effect of changing the information contained in the infall time PDFs. In both cases we used the same subsample used illustratively in Fig.~\ref{banana}. First, we reconstructed our PDFs binning only along the $R$ direction in the $(R,V)$ plane, effectively ignoring the velocity information and emulating the scenario where robust redshifts for cluster members are unavailable. The fit using this modified PDF is broadly similar to the one using the PDF including the velocity information. The preferred $\tau$ drops from $0.60^{+0.45}_{-0.45}$ to $0.00^{+0.15}_{-0.00}$~Gyr, and $\Delta t$ increases from $5.15^{+0.73}_{-0.70}$ to $5.74^{+0.14}_{-0.35}$~Gyr. The values are consistent within the quoted confidence intervals, and we note that, though $\Delta t$ and $\tau$ vary individually, the combined timescale $t_{1/2}$ increases by only $0.18$~Gyr. The statistical uncertainties on all parameters are somewhat narrower when the velocity information is not used, which at first seems surprising. However, the maximum likelihood drops from $-2702$ to $-2721$, a formally very significant ($\sim5.3\sigma$) difference. The narrower confidence intervals are a natural consequence of the poorer fit: the $\chi^2$ is larger, so the change in a parameter required to produce a given change in $\chi^2$ shrinks, apparently leading to narrower confidence intervals, but this is an illusion due to a poorer model fit.

The second test we performed was to reconstruct the infall time PDFs by dividing the $(R,V)$ plane into $50$ bins in each direction (our fiducial PDFs use $100 \times 100$ bins). In this case we recover a formally somewhat better fit, with the likelihood increasing from $-2702$ to $-2697$ ($\sim2.2\sigma$ significant). The best-fitting parameters are consistent within the confidence intervals; $\tau$ drops to $0.00^{+0.45}_{-0.00}$~Gyr and $\Delta t$ increases to $5.76^{+0.12}_{-0.36}$~Gyr, again highlighting the degeneracy between the two parameters. For this reason we prefer to focus on the `combined' timescale $t_{1/2}$, but we note that our conclusion that $\tau$ prefers small values $\lesssim 2$~Gyr appears to be robust.

\subsection{Comparison with other works}\label{subsec-compare}

Our timescales are not directly comparable to many previous studies of satellite quenching for two reasons. First, the radius at which `infall' is defined ($2.5\,r_{\rm vir}$) is larger than most previous studies which typically adopt $1.0\,r_{\rm vir}$ (with varying definitions of `virial'). As noted above, we chose this large radius to avoid the ambiguity of tracking `backsplash' subhaloes which would otherwise exit and re-enter the virial radius. A correction for this difference is relatively straightforward, since the time for a typical subhalo to move from $2.5\,r_{\rm vir}$ to $1.0\,r_{\rm vir}$ is $\sim 3$~Gyr (in detail this depends on which virial definitions are assumed). Second, some previous studies define the time for quenching since the \emph{first} time a subhalo falls into a larger halo of \emph{any} mass. Thus, for example, $30$~per~cent of satellites falling into a $\sim 10^{14} \Msun$ cluster halo had already become satellites of a lower mass group that then fell into the cluster-mass halo. Their quenching time therefore includes the time a satellite spent being `pre-processed'. In contrast, our methodology compares a quenched population ($f_{\rm passive, in}$) with an infalling population that is already `pre-processed' ($f_{\rm passive, out}$) and so isolates the quenching that is due only to falling into the current $\sim 10^{14} \Msun$ host halo. Consequently, due to the different definitions, if the infall radii were the same, our times since infall would always be shorter.

\begin{figure*}
\leavevmode \epsfxsize=2\columnwidth \epsfbox{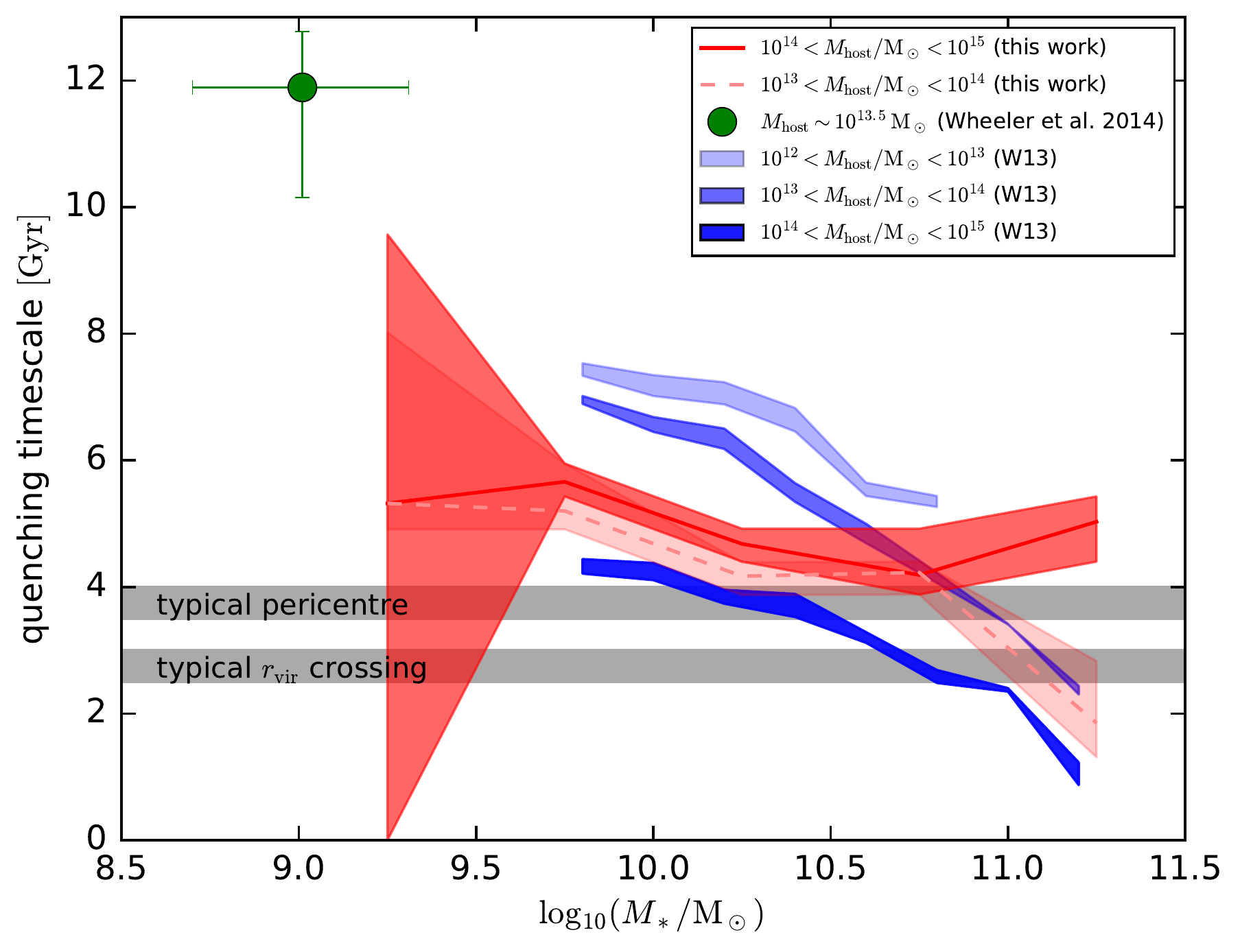}
\caption{Comparison of our combined timescale $t_{1/2}$ values (lines) and $68$~per~cent confidence intervals (shaded regions) as a function of $M_*$ with those of W13 (shaded regions represent $68$~per~cent confidence intervals) and \citet[][horizontal errorbars representing the interquartile range of $M_*$ for their sample, vertical error bars representing the uncertainty on satellite quenched fraction from 25 to 55~per~cent]{2014MNRAS.442.1396W}. Host mass ranges or representative values are as shown in the legend. In both cases, their definitions of infall time differ from ours, but we attempt to correct for the differences. See \S\ref{sec-conclusions} for the details of these corrections.\label{fig_compare}}
\end{figure*}

In Fig.~\ref{fig_compare} we show a comparison of our results with those of W13. We compare our combined timescale $t_{1/2}$ with their $t_Q$ parameter, which is similarly a combination of a delay and a transition timescale (though in W13 the `transition' time refers to the time for an individual galaxy to `fade' from blue to red). In order to compare quantitatively -- W13 uses a very different methodology to ours, but a sample with overlapping mass cuts -- we have attempted to make a correction for the different definitions mentioned above. We correct for the offset between first infall (their preferred definition) and recent infall (which corresponds to crossing $r_{\rm 200b}$), which depends on the host mass\footnote{The difference between first infall and most recent infall also depends on $M_*$, but this is a much weaker effect (Wetzel, private communication) that we neglect here.}, using the data from their fig.~2. We also correct for the median time between crossing $2.5\,r_{\rm vir}$ (i.e. our infall time) and crossing $r_{\rm 200b}$, which is $2.5$~Gyr. We make a further small correction for `ejected' (i.e. `backsplash') haloes using the timescales in \citet{2014MNRAS.439.2687W}. The total offsets we apply to the W13 results for their three host mass bins (low to high) are $2.2$, $1.1$ and $-0.1$~Gyr. The comparison is shown in Fig.~\ref{fig_compare}. We find the same trend of a decreasing quenching timescale with increasing $M_*$, though it appears the slope in our results is somewhat shallower. We also find a much weaker trend than W13, perhaps no trend, with $M_{\rm host}$ in the range probed by our sample. Some of the difference may be explained by the different treatment of `pre-processing'. At the high $M_*$, high $M_{\rm host}$ end, W13 find quenching times that, in our interpretation, correspond to quenching over a Gyr \emph{before} first entering $r_{\rm vir}$. This seems likely to be the signature of pre-processing in another group or cluster. In contrast, in our methodology which treats `pre-processed' galaxies simply as part of the passive portion of the infalling galaxy population, and so isolates the effect of the final host, quenching times are restricted to around or after the time of the first pericentric passage (marked with a horizontal gray band in the Fig.~\ref{fig_compare}).

We also plot for comparison in Fig.~\ref{fig_compare} the result of \citet{2014MNRAS.442.1396W}\footnote{We use the result as reported by \citet{2015arXiv150306803F}, which includes uncertainty estimates.}. Again, the values are not directly comparable with our own, so we attempt to adjust them to match our definitions. We increase their reported timescale by $1.7$~Gyr to account for the difference between the infall time into any more massive host and the most recent infall into a more massive host, again guided by fig.~2 of W13 (the median host mass of the \citealt{2014MNRAS.442.1396W} sample is $10^{13.5}\Msun$), and a further offset of $2.5$~Gyr to account for the travel time between $2.5\,r_{\rm vir}$ and $r_{\rm vir}$\footnote{\citet{2014MNRAS.442.1396W} define infall based on FoF group membership. This definition is unfortunately awkward for comparison; we simply assume that the edge of the FoF group corresponds to $\sim r_{\rm vir}$.}. We omit the other results reported in \citet{2015arXiv150306803F} from our comparison figure as they have no overlap in either $M_{\rm host}$ or $M_*$ with our sample.

Recently, a number of authors \citep{2014MNRAS.442.1396W, WetTolWei15, 2015arXiv150306803F, 2016MNRAS.455.2323M} have suggested that galaxies with $M \sim 10^{9} \Msun$ are significantly more resistant to quenching than satellites with both higher and lower masses. Our results for our lowest $M_*$ bin are highly uncertain, but we do seem to find an increase toward $M_* \sim 10^9$, even though we cannot make any strong statements about the timescale at this mass scale. However, the timescales we find at higher masses are clearly lower than the $11.9$~Gyr (value estimated assuming our definitions) found by \citet{2014MNRAS.442.1396W}. \citet{WetTolWei15} infer a timescale of $\sim 8$~Gyr for somewhat less massive $M \sim 10^{8.5} \Msun$ satellites of the Milky Way and M~31, and \citet{2015arXiv150306803F} find substantially lower timescales, again around smaller hosts than those in our sample. This suggests that our results are plausibly consistent with the conclusion that $M_* \sim 10^{9} \Msun$ satellites are most resistant to quenching, but that the host halo mass dependence remains to be better understood.

\subsection{Disentangling the physical mechanisms responsible for quenching}

It is interesting to consider what the observed timescale of quenching and its dependence on stellar mass reveals regarding the astrophysical mechanisms responsible for quenching star formation. A number of physical processes occur when a galaxy falls into a cluster halo. First, the accretion of dark matter and gas onto the halo is cut off while the satellite halo is still outside the virial radius. Second, ram pressure stripping may strip the hot gas halo as well as the cold gas from the disc. Finally, outflows due to galactic winds may deplete the gas that is available for star formation.

It has long been assumed in models of galaxy formation that a galaxy stops accreting gas onto its own halo when it becomes a satellite \citep[e.g.\ ][]{KauWhiGui93, ColAraFre94}, and this is also observed in SPH simulations \citep{KerKatFar09}. The cluster-centric radius at which this cut off occurs is not well known. For example, \citet[][see their fig.~8]{BahMcCBal13} argue that satellite dark matter halos stop growing within $\sim 2 r_{\rm 200c}$ of a cluster, but that their hot gas content is already reduced as far out as $5 r_{\rm 200c}$. \citet{BehWecLu14} have also shown that dark matter accretion ends roughly when the satellite is as far out as $\sim 2 r_{\rm vir}$. So it is likely that the cut-off of accreting gas occurs further out than the fiducial virial radius, and closer to our `backsplash' limit of 2.5 $r_{\rm vir}$.

Even if the supply of new gas is cut off, the existing reservoir of cold and hot gas is large enough to sustain star formation in excess of the Hubble time at typical star formation rates: $t \sim (M_{\rm baryon} - M_{*})/{\rm SFR}(M_{*})$. Dividing both numerator and denominator by $M_{*}$ gives $t \sim (f_{\rm baryon}/f_{*}-1)/{\rm SSFR}$, where the fraction of mass in baryons $f_{\rm baryon} \sim 0.15$ and the fraction of total mass in stars $f_{*} \sim 0.01$ for the lowest stellar mass galaxies \citep{HudGilCou15}. Therefore, to explain the short quenching times, additional mechanisms are required to remove or heat the existing gas. As discussed in \S\ref{sec-introduction}, ram pressure stripping of cold gas is clearly seen in galaxy clusters. The key results of this paper are that (i) the quenching occurs approximately at or shortly after pericentre passage, (ii) after the delay $\Delta t$, it is $100$~per~cent effective, (iii) that the time for low-mass galaxies to quench is slightly longer than the time for higher mass galaxies and (iv) the infalling population transitions to become the cluster population relatively quickly (once the delay $\Delta t$ has elapsed), on a timescale $\tau \lesssim 2$~Gyr. Because ram pressure stripping is strongest close to pericentre, the observed timing of the quenching is in broad agreement with the ram pressure stripping model. However, whether the remaining two observations are in accordance with this model is less clear. Ram pressure stripping is expected to be more effective for low mass satellites because the restoring force of the disc is lower, which would argue against the model. However, larger galaxies are more affected by dynamical friction and the ram pressure is very sensitive to speed; it is proportional to the square of the speed through the intra cluster medium. Smaller satellites are likely still proportionally more affected by ram pressure \citep{BahMcC15}, so the trend remains puzzling unless either smaller satellites have some intrinsic property causing them to take longer to cease forming stars or ram pressure is not the dominant trigger of quenching for satellites of all masses in clusters, or some combination of both.

\begin{figure}
\leavevmode \epsfxsize=\columnwidth \epsfbox{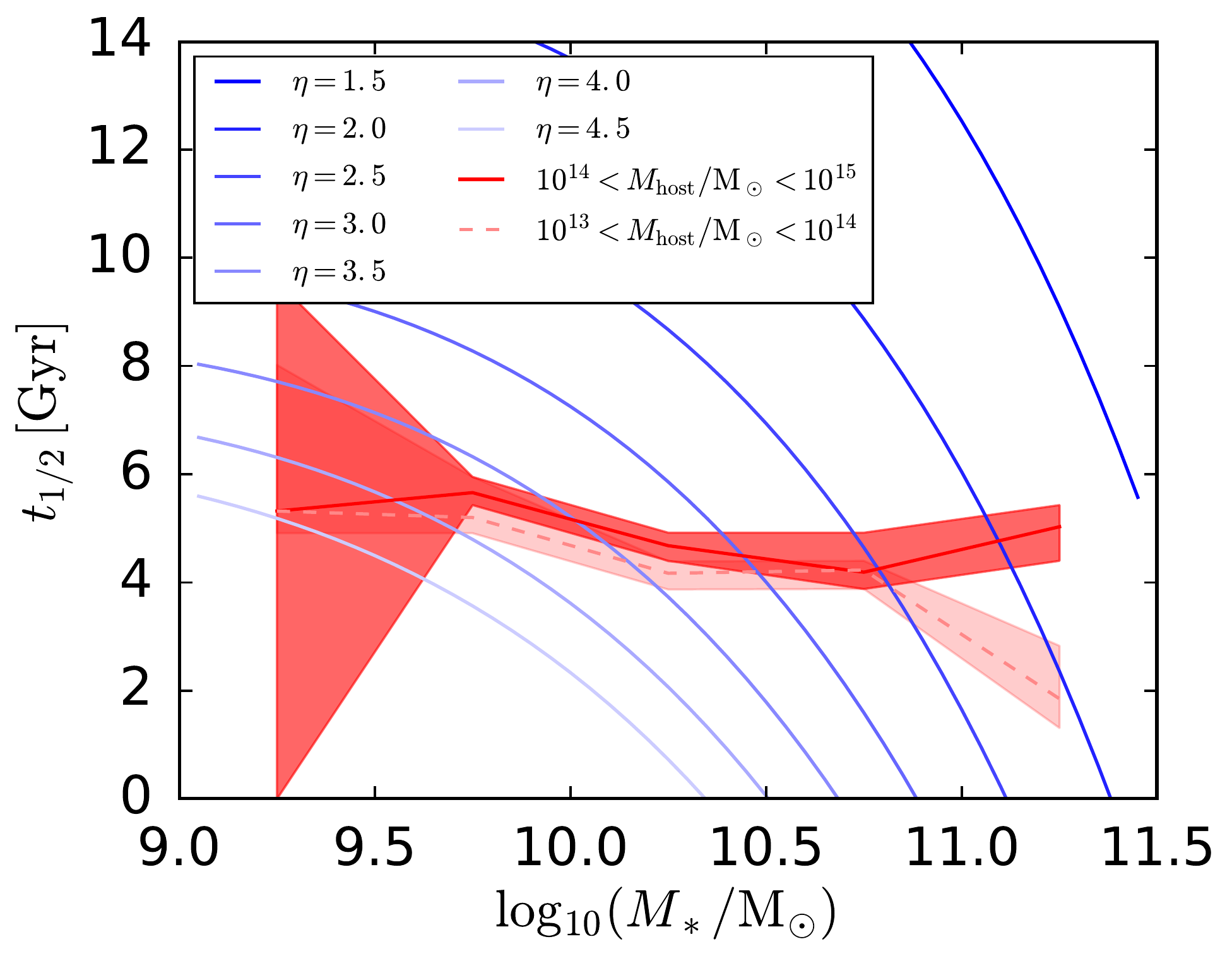}
\caption{Comparison of our quenching timescales as a function of $M_*$ with the simple overconsumption models of \citet{McGBowBal14}, parametrized by the mass-loading factor $\eta$.\label{fig_mcgee}}
\end{figure}

Recently, \citet{McGBowBal14} have advocated for the combination of a cut-off in gas accretion as a galaxy falls into a cluster, coupled with strong outflows driven by galactic winds (a model which they dub `overconsumption'), to explain the quenching timescales. In their model, the key parameter is the mass-loading of the winds: $\eta = \dot{M}_{\rm out}/{\rm SFR}$, where $\dot{M}_{\rm out}$ represents the rate at which gas is permanently ejected from the satellite's halo.  The quenching time is highly sensitive to $\eta$: if $\eta$ is too low then the quenching time is longer than the Hubble time, too high and the quenching time rapidly approaches zero. \citet{McGBowBal14} and \citet{BalMcGMok15} argue that  $\eta \sim 1.5$, independent of mass.  We have used eq.~7 of \citet{McGBowBal14} to fit their model to our quenching times, assuming no stripping. We have adopted the stellar-to-halo mass relation from weak lensing \citep{HudGilCou15}, and the low redshift SFR of \citet{2007ApJS..173..267S}. Finally, we have also assumed that the `clock' for (over)consumption starts ticking when the satellites crosses $2.5 r_{\rm vir}$. The predictions for contours of constant $\eta$ are shown in Fig.~\ref{fig_mcgee} . The data are fit with a slow varying $\eta$ that ranges from $2.0$ at high mass to $4.0$ at low mass.

A slowly-varying $\eta$ model is in conflict, however, with other results on the mass-loading of outflows. In particular, one would expect galaxies with shallower potential wells to have more efficient outflows. This is found in numerical simulations: \citet{MurKerFau15} find that the mass-loading factor scales as $\eta \propto v_{\rm circ}^{-1}$, and \citet{2016arXiv160408244K} finds a constant $\eta\sim 8$ for low mass systems, decreasing with a power law slope of about $-1.8$ above $M_* = 10^{10}\Msun$. From analytic arguments based on the scaling relations of low mass galaxies and the baryonic TF relation, \citet{2012MNRAS.424.3123D} finds that $\eta \sim v_{\rm circ}^{-2}$. These three models would predict higher values of $\eta$ at low stellar mass, and hence short quenching times for low mass galaxies for which winds efficiently remove the gas.

The above derived $\eta$ is an upper limit on the true $\eta$, because ram pressure stripping of the hot gas reservoir will substantially reduce the amount of gas potentially available for star formation \citep{BahMcC15}.  Furthermore, the `effective' $\eta$ is likely to be higher for infalling satellites than similar counterparts in the field, because weak outflows that in field  galaxies would return as a galactic fountain are instead stripped by ram pressure \citep{BahMcC15}.

Overall, the timing of quenching near pericentre suggests that ram pressure stripping plays a role. However, the fact that low-mass galaxies have longer quenching time delays than high mass galaxies is difficult to understand, because whatever the mechanism of gas removal, whether ram pressure stripping or galactic winds, it should be more effective in low-mass galaxies with shallower potential wells.

\section{Summary and outlook}\label{sec-summary}

We have compared subhalo orbit libraries in projected phase space to star formation rates of SDSS galaxies. This method isolates the environmental effects of the most recent host; in this paper, this is a cluster of mass $10^{13}-10^{15}\Msun$. The key results of this paper are as follows:
\begin{enumerate}
\item Quenching occurs after a delay time $\Delta t$, measured from first crossing of $2.5r_{\rm vir}$. This delay time is typically $3.5$--$5$~Gyr, with higher mass galaxies quenching slightly earlier. In most cases, this corresponds to times near or shortly after first pericentric approach. All galaxies are quenched on first infall, and before apocentre.
\item The delay time does not depend (or depends very weakly) on the host halo mass, over the relatively narrow range probed by our sample.
\item Once quenching begins, the timescale $\tau$ for the galaxy population to transition from resembling the galaxies outside the cluster (described by $f_{\rm passive, out}$) to those processed by the cluster (described by $f_{\rm passive, in}$) is fairly short, $\lesssim 2$~Gyr, and usually consistent with $0$~Gyr. Note that this timescale is distinct from the timescale for individual galaxies to transition from an active to a passive state, i.e. the timescale for `crossing the green valley'.
\item After the delay has elapsed, the quenching is $100$~per~cent effective, i.e.~all active galaxies that fell in longer ago than $\approx\Delta t + 2\tau$ are passive. The observed fraction of star-forming galaxies in rich clusters is therefore due to a combination of interlopers and galaxies that are falling in for the first time.
\end{enumerate}

These results appear to be in reasonable agreement with some previous work \citep[W13,][]{2014MNRAS.442.1396W}, after correction the fact that delay times in these works are measured at first accretion and at a different radius.

In this paper, we have shown how an orbit library can be used to deproject infalling, backsplash and virialized populations. Here we have compared our projected models with SFR data in PPS, but only for a very simple parameterization of the SFR distribution; this could be extended to leverage the additional information contained in the full SFR distributions. Furthermore, there is no reason to limit the comparison to only SFR, particularly since it is well know that morphology is also correlated with environment. Tidal or harassment effects may also affect the \emph{structures} of discs, possibly stripping them (reduction in stellar mass and radius) or puffing them up so that they are identified morphologically as bulges. In addition, here we have limited ourselves to the correlation between time since infall and PPS position, however much more information is contained in orbit libraries. As redshift surveys continue to improve and grow, we expect that increasingly subtle effects can be teased out of the data.

\section*{Acknowledgements}\label{sec-awknowledgements}

We wish to thank Michael Balogh, Coral Wheeler, Laura Sales and Andrew Wetzel for useful discussions. We thank Sara Ellison for providing tabulated stellar mass and SFR data and Andrew Wetzel for providing results in electronic form. We also thank the authors of the Bolshoi and MDR1 simulations for making their simulation outputs publicly available, and Peter Behroozi for assistance with the halo finder and merger tree codes. MJH acknowledges support from an NSERC Discovery grant. This research has made use of NASA's Astrophysics Data System.

\bibliography{paper}

\label{lastpage}

\end{document}

%% file: figs/table.tex
\begin{tabular}{llllllll}
  \hline
  $M_{\rm host}\,[{\rm M}_\odot]$ & $M_*\,[{\rm M}_\odot]$ & $f_{\rm passive, in}$ & $f_{\rm passive, out}$   & $\Delta t\,[{\rm Gyr}]$  & $\tau\,[{\rm Gyr}]$    & $t_{1/2}\,[{\rm Gyr}]$ &${\rm max}(\log_e\mathcal{L})$  \\
  \hline
                                      & $10^{9}-10^{9.5}$                         & $0.53^{+0.47}_{-0.00}$   & $0.12^{+0.02}_{-0.01}$  & $5.52^{+0.00}_{-5.52}$    & $0.00^{+11.00}_{-0.00}$  & $5.32^{+2.70}_{-0.41}$ & $-237$                     \\
                                      & $10^{9.5}-10^{10}$                        & $0.84^{+0.16}_{-0.10}$   & $0.21^{+0.02}_{-0.03}$  & $5.20^{+0.00}_{-5.20}$    & $0.00^{+1.50}_{-0.00}$   & $5.20^{+0.75}_{-0.29}$ &$-927$                     \\
  $10^{13}-10^{14}$                     & $10^{10}-10^{10.5}$                       & $0.96^{+0.04}_{-0.10}$   & $0.51^{+0.02}_{-0.02}$  & $3.07^{+1.03}_{-0.39}$    & $1.80^{+0.30}_{-1.80}$   & $4.17^{+0.22}_{-0.30}$ & $-2439$                    \\
                                      & $10^{10.5}-10^{11}$                       & $0.99^{+0.01}_{-0.06}$   & $0.70^{+0.02}_{-0.02}$  & $4.33^{+0.00}_{-1.88}$    & $0.00^{+1.65}_{-0.00}$   & $4.23^{+0.17}_{-0.35}$ &$-2032$                     \\
                                      & $10^{11}-10^{11.5}$                       & $0.94^{+0.06}_{-0.04}$   & $0.84^{+0.01}_{-0.04}$  & $1.91^{+0.09}_{-1.91}$    & $0.00^{+1.75}_{-0.00}$   & $1.85^{+0.98}_{-0.54}$ &$-260$                      \\
  \hline
                                      & $10^{9}-10^{9.5}$                         & $0.63^{+0.37}_{-0.04}$   & $0.20^{+0.01}_{-0.02}$  & $0.00^{+1.78}_{-0.00}$    & $3.00^{+4.50}_{-3.00}$   & $5.32^{+4.25}_{-5.32}$ & $-1008$                     \\
                                      & $10^{9.5}-10^{10}$                        & $0.99^{+0.01}_{-0.03}$   & $0.23^{+0.02}_{-0.01}$  & $5.15^{+0.73}_{-0.70}$    & $0.60^{+0.45}_{-0.45}$   & $5.66^{+0.29}_{-0.23}$ & $-2702$                     \\
  $10^{14}-10^{15}$                    & $10^{10}-10^{10.5}$                       & $0.99^{+0.01}_{-0.01}$   & $0.45^{+0.01}_{-0.01}$  & $3.95^{+0.15}_{-0.36}$    & $0.90^{+0.75}_{-0.15}$   & $4.68^{+0.24}_{-0.28}$ & $-6730$                      \\
                                      & $10^{10.5}-10^{11}$                       & $1.00^{+0.00}_{-0.03}$   & $0.69^{+0.02}_{-0.01}$  & $3.46^{+0.37}_{-0.33}$    & $1.50^{+0.45}_{-0.45}$   & $4.19^{+0.73}_{-0.31}$ & $-6165$                      \\
                                      & $10^{11}-10^{11.5}$                       & $1.00^{+0.00}_{-0.07}$   & $0.86^{+0.02}_{-0.02}$  & $5.10^{+0.00}_{-2.43}$    & $0.00^{+2.50}_{-0.00}$   & $5.03^{+0.40}_{-0.63}$ & $-758$                       \\
  
\hline
\end{tabular}